\documentclass{ws-mpla}

\title{Kinematic equivalence between models driven by DBI field with constant $\gamma$ and exotic holographic quintessence cosmological models}
\author{M\'onica Forte}

\usepackage[super]{cite}
\usepackage{graphicx}
\RequirePackage[T1]{fontenc}
\RequirePackage{fix-cm}
\RequirePackage{graphicx}
\RequirePackage{mathptmx} 
\RequirePackage{latexsym}
\RequirePackage{flushend}
\RequirePackage[numbers,sort&compress]{natbib}
\RequirePackage[colorlinks,citecolor=blue,urlcolor=blue,linkcolor=blue]{hyperref}
\RequirePackage{amsmath}

\newcommand{\ben}{\begin{eqnarray}}
\newcommand{\een}{\end{eqnarray}}
\newcommand{\be}{\begin{equation}}
\newcommand{\ee}{\end{equation}}
\newcommand{\n}{\label}
\newcommand{\no}{\noindent}

\begin{document}

\title{Kinematic equivalence between models driven by DBI field with constant $\gamma$ and exotic holographic quintessence cosmological models}


\address{{Departamento de F\'isica, Facultad de Ciencias Exactas y Naturales, Universidad de Buenos Aires, 1428 Buenos Aires, Argentina}\\
forte.monica@gmail.com}

\maketitle

\pub{Received (Day Month Year)}{Revised (Day Month Year)}

\begin{abstract}
We show the kinematic equivalence between cosmological models driven by Dirac-Born-Infeld fields $\phi$ with constant proper velocity of the brane and exponential potential $V=V_0e^{-B\phi}$ and interactive cosmological systems with Modified Holographic Ricci type fluids as dark energy in flat Friedmann-Robertson-Walker cosmologies.

\keywords{DBI field; interacting dark energy; Modified Holographic Ricci.}
\end{abstract}

\ccode{04.20.-q,04.60.Cf,98.80.-k,95.36.+x}

\section{Introduction}	

The most successful explanation for the origin of the observed temperature fluctuations of the cosmic microwave background (CMB) (\cite{Linde:2005ht},\cite{Lyth:1998xn}) is the inflationary paradigm, which however, so far have not found a satisfactory theoretical foundation. The  favorite candidate to explain these observations is the String Theory \cite{Baumann:2009ni} and inside it the very interesting model of Dirac-Born-Infeld (DBI) inflation \cite{Silverstein:2003hf,Alishahiha:2004eh,Chen:2004gc,Chen:2005ad,Shandera:2006ax}. In DBI models the inflation is driven by the motion of a D3-brane in a warped throat region of a compact internal space and the non-canonical DBI field $\phi$ is interpreted as the position of the D-brane. The DBI action includes a potential arising from the quantum interaction between D-branes and the brane tension $\mathcal{T}$ encodes geometrical information about the throat region of the compact space. 
On the other hand, related to the amount of dark energy (DE) required to generate the negative pressure justifying the observed expansion of our universe, appears the holographic principle stated as ``The number of degrees of freedom in a bounded system should be finite and is related to the area of its boundary"  \cite{Bousso:2002ju}.
The above prescription applied to cosmology suggests that the ultraviolet (UV) cutoff scale of a system is connected to its infrared (IR) cutoff scale $\Lambda\equiv \rho_{\Lambda}^{1/4}$ \cite{Cohen:1998zx}.   When $L$ (IR length) is a very descriptive length of the system which total energy is $L^3\rho_{\Lambda}$, the application suggests that the region will not decay into a black hole if its mass not exceed the mass of a black hole of the same size. Then $L$ is such that $L^3\rho_{\Lambda}\leq LM_p^3 $ with $M_p$ the reduced Planck mass. The largest $L$ allowed is the one which saturates the above inequality and leads to a holographic dark energy proportional to $L^{-2}$. Hence, this principle connects the dark energy based on the quantum zero-point energy density caused by a short distance cutoff $\Lambda$ with an IR cutoff that is usually taken as Hubble horizon, particle horizon, event horizon \cite{Hsu:2004ri,Li:2004rb}, or generalized IR cutoff \cite{Nojiri:2005pu}. 
While many of the works with holographic fluids do not take into account the non-gravitational interaction with dark matter, the holographic dark energy is interactive because of the nature of the expression describing its density of  energy. As shown in \cite{Chimento:2011dw}, its density of  energy is affected by the other component of the system even though there is no external interaction connecting them. Holographic dark energy is self-interactive and cosmological models with an interaction between components are essential to alleviate problems such as the so-called crisis age \cite{Alcaniz:1999kr,Forte:2013fua}.  
Then, we consider a justificatory model of cosmological expansion using dark matter (DM) interacting with a holographic fluid amended as dark energy as follows. A model where a perfect fluid with constant equation of state (EoS) $\omega_{DM} $ and density of energy $\rho_{DM}$ interacts with an holographic fluid whose density of energy is $\rho_{DE}=2(\dot H+3\alpha H^2/2)/(\alpha-\beta)$ (called a modified Ricci type) is likely to be identified with a unified model driven by an exotic quintessence $\psi$ as was done in \cite{Chimento:2007da} for two arbitrary ideal fluids where both EoS are constant. This can be done because of the particular expression selected for the density of energy $\rho_{DE}$ where $ \alpha$ and $\beta$ are constants.

Our paper is organized as follows. In Section II we consider a Dirac-Born-Infeld cosmological model  with $\gamma=constant$ and obtain the corresponding differential equation for the factor of scale of the Friedmann Robertson Walker background geometry. In Section III we analyze the exotic quintessence  unified model that represents an interactive system composed by dark matter and modified Ricci type holographic fluid as dark energy, obtaining the equation that governs the evolution of the factor of scale. In Section IV we establish the linking between both models and  make conclusions.

\section{The Dirac-Born-Infeld field with $\gamma=constant$ }
\label{DBI}

In Friedmann Robertson Walker (FRW) background geometry we consider a cosmological model commanded by a non-canonical DBI field  $\phi$,  derivable from a lagrangian $\mathcal{L}$ 
\be
\n{13}
\mathcal{L}=\gamma\dot\phi^2/(\gamma+1)-V(\phi),\ \  \gamma =\frac{1}{\sqrt{1-f(\phi)\dot\phi^2}}>1,
\ee
\no where $f(\phi)= \mathcal{T}^{-1} > 0$ is the warp factor and $\sqrt{f(\phi)}\dot\phi$ may be interpreted as the proper velocity of the brane. Here we use an exponential potential $V(\phi)=V_0 e^{-B\phi}$ and $\dot\phi$ is the derivative respect the cosmic time. The customary perfect fluid interpretation $p_{\phi} = (\Gamma - 1)\rho_{\phi}$, where $p_{\phi}$ are the pressure and $\rho_{\phi}$ are the density of energy describes a barotropic index  $\Gamma$ of the DBI field and leads to 

\be
\n{11}
\rho_{\phi}=\frac{\gamma^2}{\gamma+1}\dot\phi^2 + V(\phi),
\ee
\be
\n{12}
p_{\phi}=\frac{\gamma}{\gamma+1}\dot\phi^2 - V(\phi),
\ee
and so
\be
\n{14} 
-2\dot H=\Gamma\rho = \gamma\dot\phi^2 \qquad H=d\ln{a}/dt.
\ee

 For a DBI scenario with $\gamma=\gamma_0=const.$ the equation (\ref{11}) lets obtain the equation of movement for the DBI field
\be
\n{15} 
\ddot\phi + \frac{3(\gamma_0+1)}{2\gamma_0}H\dot\phi+(\gamma_0+1) \frac{V'}{2\gamma_0^2}=0,
\ee
\no where $V'=dV/d\phi=-BV$ and which first integral is 
\be
\n{16} 
\dot\phi= \frac{B}{\gamma_0}H+\frac{b}{a^{3( \gamma_0+1)/2\gamma_0}},
\ee
\no with the constant of integration $b$.

Equations (\ref{14}) and (\ref{16}) converge to describe the ordinary second order differential equation that must be hold by the factor of scale $a$,
\be
\n{17} 
2\dot H+\frac{B^2}{\gamma_0}H^2+2\frac{bBH}{a^{3(\gamma_0+1)/2\gamma_0}}+\frac{\gamma_0b^2}{a^{3(\gamma_0+1)/\gamma_0}}=0.
\ee
\vskip0.5cm

\section{The exotic quintessence representation of interacting modified holographic Ricci type dark energy}
\label{exotic}

The Einstein equations in Friedmann-Robertson-Walker geometry for this unified model are read
\be
\n{1}
3H^2 =\rho_{DM}+\rho_{DE}=\rho,
\ee
\be
\n{2}
-2\dot H=\rho+p=(1+\omega_{DM})\rho_{DM}+(1+\omega_{DE})\rho_{DE}.
\ee
\vskip0.5cm
The particular length of the cut off used in the description of $\rho_{DE}$, as linear combination of the Hubble parameter $H$ and its time derivative $\dot H$, lets  write (\ref{2}) as
\be
\n{3}
-2\dot H=\alpha\rho_{DM}+\beta\rho_{DE}.
\ee
With (\ref{2}) and (\ref{3}) we define the scalar representation of the global interaction system through the time derivative of the quintessence field $\psi$ as
\be
\n{4}
-2\dot H=\dot\psi^2.
\ee

From (\ref{3}) and (\ref{4}), the equation of movement for the scalar field $\psi$ is
\be
\n{5}
\ddot\psi+\frac{3}{2}\alpha H \dot\psi +\frac{(\alpha-\beta)}{2}\frac{\dot\rho_{DE}}{\dot\psi}=0
\ee
and the functional form of $\dot\rho_{DE}$ complete the definition of the representation. The most convenient option is to consider $\rho_{DE}$ as an exponential potential, that is
\be
\n{6}
\dot\rho_{DE}+\nu\dot\psi\rho_{DE}=0
\ee
\no with the constant $\nu $, so that $\rho_{DE}=\mathcal{U}(\psi)=\mathcal{U}_0e^{-\nu\psi}$.

The ``exotic" nickname for this scalar field is understood noting that its density of energy and pressure
\be
\n{7}
\rho=\frac{\dot\psi^2}{\alpha}+\frac{(\alpha-\beta)}{\alpha}\ \mathcal{U}(\psi),
\ee
\be
\n{8}
p=\frac{(\alpha-1)}{\alpha}\dot\psi^2-\frac{(\alpha-\beta)}{\alpha}\ \mathcal{U}(\psi),
\ee
\no are reduced to a usual quintessence for $\alpha=2$ and $\beta=0$.  

For the option (\ref{6}), the first integral of the equation (\ref{5}) is
\be
\n{9}
\dot\psi=\nu H + \frac{\mu}{a^{3\alpha/2}},
\ee
\no where $\mu$ is a constant of integration, and the equation (\ref{4}) lets  write the second-order differential equation for the factor of scale $a$ as
\be
\n{10}
2\dot H+\nu^2H^2+\frac{2\nu\mu H}{a^{3\alpha/2}}+\frac{\mu^2}{a^{3\alpha}}=0.
\ee

\section{Conclusions}

Equations (\ref{17}) and (\ref{10}) are formally similar and making the identification 
$$\alpha=\frac{\gamma_0+1}{\gamma_0}, \quad \nu=\frac{B}{\sqrt{\gamma_0}}, \quad \mu=\sqrt{\gamma_0}b,$$
$$\psi=\sqrt{\gamma_0}\phi, \qquad V_0=\frac{(\alpha-\beta)}{\alpha} \mathcal{U}_0,$$
between the fields and the parameters, they become the same. 

We conclude that both models, one driven by a DBI field with a constant measure of the ``relativistic" motion $\gamma_0$ and exponential potential and the other, driven by an exotic quintessence with exponential potential, are described by the same factor of scale and so they are geometrically equivalent. 
Note that the kinematical equivalence between usual quintessence models driven by an exponential potential and cosmologies driven by  k-essences with kinetic function $F(x)= 1+mx$ and potential of inverse quadratic type has been proved in Bianchi I type and FRW metrics. But here, the equivalence obtained connects an ``exotic" holographic quintessence as unified model of DM interacting with modified holographic Ricci type DE and DBI  cosmologies.

As stated Liddle et al in \cite{Gao:2009me}, the particular type of quintessence that we nicknamed exotics \cite{Chimento:2007da} suffers from the problem of not having a lagrangian from which to derive the model. A possible solution to this lack comes from the equivalence shown above. The link with the DBI description suggests that there is a functional relationship between the time derivative of the field and the field itself, set by the constancy of the speed of the brane itself.
In the case of the exotic holographic quintessences $\psi$ and with the above idea in mind we propose the existence of a relationship $\dot\psi(\psi)$ between the time derivative of the field and the field itself, described through an arbitrary function $g(\psi)$ such that $g(\psi)\dot\psi^2=m_0^2$, with $m_0=const$. Also we assume that $g(\psi)$ and $\alpha$ parameter are related through

$$\alpha=\frac{2g\dot\psi^2}{1+g\dot\psi^2}.$$

With these requirements it is easy to show that the exotic holographic quintessence models are derivable from lagrangians of type
$$\mathcal{L}_{EHQ}=\frac{g\dot\psi^2-1}{2g}-U(\psi)\ \ \textrm{with}\ \ g\dot\psi^2=\frac{\alpha}{2-\alpha}.$$

Moreover,  when solutions for $a(t)$ and $\phi(t)$ are obtained as in \cite{Chimento:2010un}, we can relate the functional form of the tension on D-brane $\mathcal{T}$ as it was generated by an interactive cosmological system. For example, the power law solutions $a= t^{2\gamma_0/B^2}$ and $\phi=2\ln t/B$ are invertible and determine that the tension on the brane  has the expression $$\mathcal{T}=4\frac{(\alpha-1)}{\nu^2\alpha(2-\alpha)}e^{-\nu\psi}=4\frac{(\alpha-1)}{\nu^2\alpha(2-\alpha)}e^{-B\phi}$$  where $\alpha,\nu$ and $\psi$ 
are characteristics of the quintessential representation of the equivalent interactive system.

In some sense we can say that the interactive system ``calibrates" the tension on the D-brane and conversely, the condition $\mathcal{T}>0$ sets limits to the possible values of $\alpha$ that should belong to the interval $1<\alpha<2$.

\end{document}